\documentclass[aps,prb,twocolumn,superscriptaddress,amsmath,amsfonts,showpacs]{revtex4}

\usepackage{graphicx}
\usepackage{bm} 
\usepackage{braket}
\usepackage[usenames]{color}
\usepackage{comment}

\begin{document}


\title{Index theorem for topological heterostructure systems}


\author{Ken Shiozaki}
\affiliation{Department of Physics, Kyoto University, Kyoto 606-8502, Japan}
\author{Takahiro Fukui}
\affiliation{Department of Physics, Ibaraki University, Mito 310-8512, Japan}
\author{Satoshi Fujimoto}
\affiliation{Department of Physics, Kyoto University, Kyoto 606-8502, Japan}




\date{\today}

\begin{abstract}
We apply the Niemi-Semenoff 
index theorem 
to an $s$-wave superconductor junction system attached with a magnetic insulator on the surface of a three-dimensional topological insulator. 
We find that the total number of the Majorana zero energy bound states is governed 
not only by the gapless helical mode 
but also by the massive modes localized at the junction interface. 
The result implies that the topological protection for Majorana zero modes in class D heterostructure junctions may be broken down
under a particular but realistic condition.

\end{abstract}

\pacs{73.20.-r, 71.10.Pm, 74.45.+c}


\maketitle



\section{Introduction}

Zero energy bound states in vortex cores of superconductors have been of much current interest in condensed matter physics. 
Some classes of the vortex Majorana states, obeying non-Abelian statistics, 
may serve as qubits for quantum computation. \cite{Kitaev,Nayak,Goldstein,BWT}
There are various theoretical proposals for realizing non-Abelian Majorana fermions in the core of vortices in topological superconductors,
e.g., a chiral $p$-wave superconductor, etc.\cite{RG,MR,Sau,Ivanov,FK,MTF,Alicea,TK2,Sato,Santos}
Besides such vortex zero modes, topological superconductor junction systems, in which the order parameter changes sharply in real space, possess generically
non-Abelian Majorana fermions.\cite{FK} 
A useful method for charactering the existence of the zero energy bound states localized at point defects such as vortices or point intersections consisting of the junction interfaces 
is the index theorem for an open infinite space,
derived by Callias \cite{Callias} and by Weinberg \cite{Weinberg}, and generalized by Niemi-Semenoff. \cite{NS1} 
This theorem reveals the relationship between the zero energy bound states and the topology of background fields at large distance from the point defects. 
In this paper, we investigate the index theorem for the heterostructure system involving the topological insulator (TI), mainly focusing on the superconductor-TI-ferromagnet insulator junctions. 
We find that the number of Majorana zero modes is controlled not only by the phase winding of the superconducting gap, but also by non-topological massive bound states localized at the junctions. 

The organization of this paper is as follows.
In Sec.\ref{sec2}, we first present our main results for the index of the superconductor-TI-ferromagnet insulator junctions, 
and discuss its physical implications.
We give, in particular, a physical explanation on how non-topological massive bound states
affect the index for Majorana zero energy modes. 
Our results are based on the celebrated Niemi-Semenoff index theorem.
To make this paper self-contained, 
we briefly review the Niemi-Semenoff index theorem in Sec.\ref{sec3}. 
In Sec.\ref{sec4}, we apply the Niemi-Semenoff index theorem to the superconductor-TI-ferromagnet insulator junctions, 
and obtain the index theorem for topological heterostructure systems. 
In Sec.\ref{sec5}, we also apply our results to
topological insulator-ferromagnetic insulator heterostructure systems. 
We conclude in Sec.\ref{sec6} with some discussions.

\section{Setup and main results}
\label{sec2}

\begin{figure}[!b]
 \begin{center}
  \includegraphics[width=\linewidth, trim=0 0cm 0 0cm]{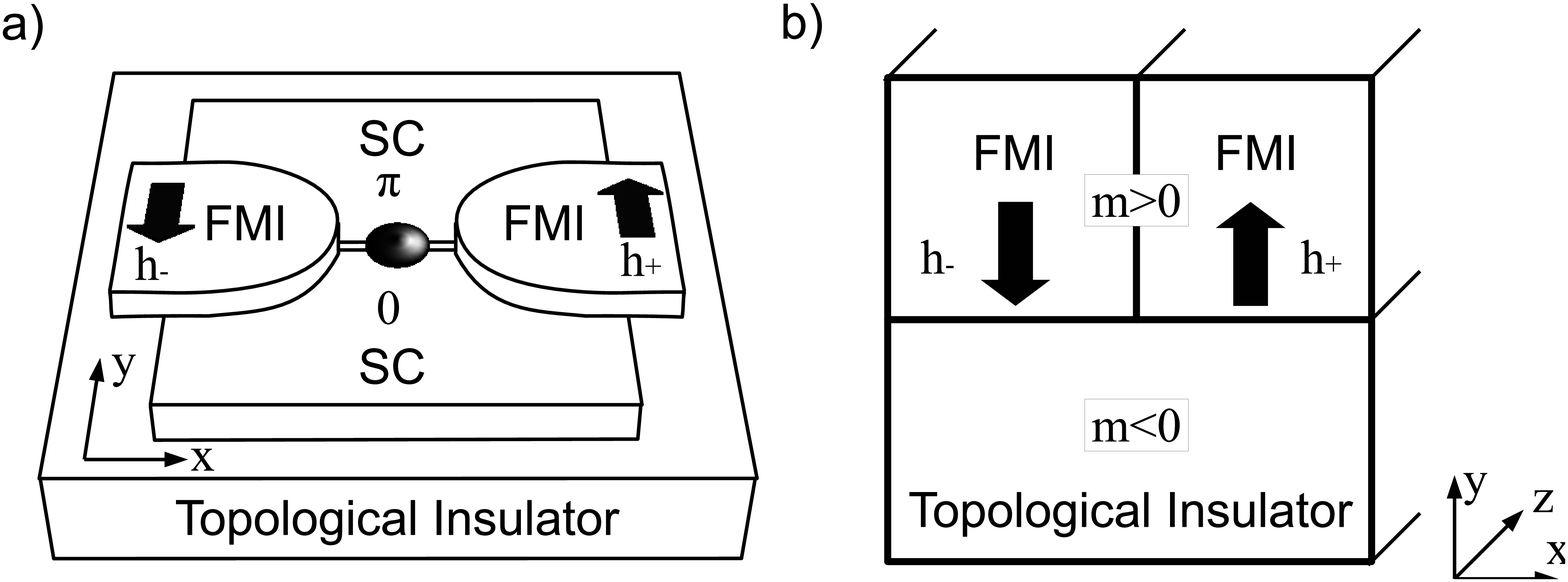}
 \end{center}
 \caption{(a) The heterostructure geometry for an $s$-wave superconductor (SC) $\pi$-junction and a ferromagnetic insulator (FMI) on the surface of a topological insulator. 
 A filled circle at the interface is a point defect formed by the intersection of the $\pi$-junction and the ferromagnetic domain wall.
 (b) The heterostructure geometry for a topological insulator-ferromagnetic insulator tri-junction. }
 \label{fig1}
\end{figure}

We consider the heterostructure system composed of an $s$-wave superconductor $\pi$-junction and ferromagnetic insulators
placed on a TI, as depicted in Fig. \ref{fig1}(a), and investigate the zero energy bound state localized at a point-like defect formed by the intersection
of the $\pi$-junction interface and the ferromagnetic domain wall. 
The effective Bogoliubov-de Gennes Hamiltonian is written as 
\begin{equation}
\begin{split}
\mathcal{H} &= -i v \tau_3 \sigma_j \partial_j + \Delta_1 \tau_1 + \Delta_2 \tau_2 + \bm{h} \cdot \bm{\sigma} -\mu \tau_3,  \\
\end{split}
\label{eq11.5}
\end{equation}
where $j = 1,2 $, $\bm{\tau}= (\tau_1, \tau_2, \tau_3)$ and $\bm{\sigma} = (\sigma_1,\sigma_2, \sigma_3)$ are the Pauli matrices for the Nambu and spin space respectively, 
$v$ is the velocity of the Dirac fermion, 
$\Delta_1$ and $\Delta_2$ are the real and imaginary parts of the gap function, $\bm{h} \cdot \bm{\sigma}$ is the Zeeman term, and $\mu$ is the chemical potential. 
Note that for the $\pi$-junction considered here, $\Delta_2=0$.
It is also assumed that the thickness of the superconducting film is sufficiently smaller than the penetration depth, 
and hence $z$-dependence of $\bm{h}$ is negligible.
This system (\ref{eq11.5}) belongs to class D in the Altland-Zirnbauer symmetry classes\cite{Schnyder,Kitaev2}, 
and the vortex zero modes obeying non-Abelian statistics \cite{FK} are classified as the $\mathbb{Z}_2$ invariant.
The index theorem which is a main tool in this paper is applicable only to systems with chiral symmetry, 
i.e. $\Pi\mathcal{H}\Pi^{\dag}=-\mathcal{H}$ is satisfied for a unitary operator $\Pi$. 
This symmetry is, however, not preserved for (\ref{eq11.5}), because of $h_3\sigma_3$ and $-\mu\tau_3$ terms. 
Nevertheless, as will be clarified below, the $\mathbb{Z}_2$ invariant of (\ref{eq11.5}) can be generically obtained from the index calculated for the case with $h_3=\mu=0$. 
Thus, we first neglect these two terms to calculate the index.
We furthermore omit the $h_1\sigma_1$ term to simplify the analysis, 
since this term does not affect the index of our system, as long as
$h_1$ is sufficiently small, and does not close the bulk energy gap.
Then, the Hamiltonian is reduces to that 
with chiral symmetry (class BDI), $\tau_3 \sigma_3 \mathcal{H} \tau_3 \sigma_3 = -\mathcal{H}$, 
\begin{equation}
\begin{split}
\mathcal{H} &= -i v \tau_3 \sigma_j \partial_j  + \Delta_1 \tau_1 
+ h_2 \sigma_2, \\
\end{split}
\label{eq12}
\end{equation}
and its ground state is classified by $\mathbb{Z}$. This enhanced topological number can be computed by the index theorem described in detail in the
next section. 
Note here that particle-hole symmetry, $\tau_2 \sigma_2 \mathcal{H}^* \tau_2 \sigma_2 = -\mathcal{H}$, valid for (\ref{eq11.5}) as well as (\ref{eq12}),
ensures that the number of vortex zero modes are conserved modulo 2 even if the 
neglected chiral-symmetry-breaking terms are switched on again. Thus, the $\mathbb{Z}_2$ invariant of (\ref{eq11.5}) 
can be derived from
the parity of the index of (\ref{eq12}). 

\begin{figure}[!h]
 \begin{center}
  \includegraphics[width=\linewidth]{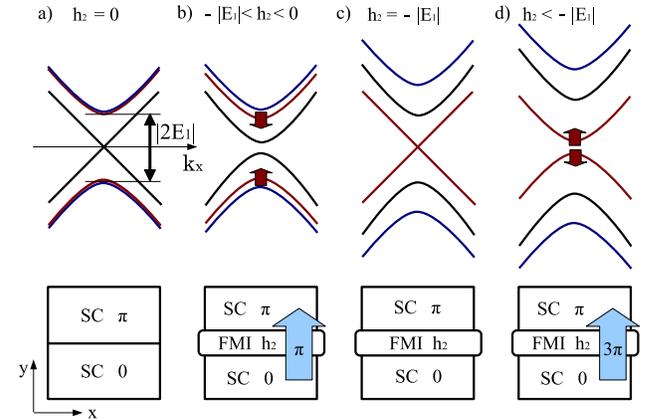}
 \end{center}
 \caption{(Color online) A schematic picture of the band inversion of massive bound states.
 (a) Black lines denote the energy band of helical Majorana fermion, while red and blue lines denotes that of massive bound states. 
 Lower panels indicate the right-half part of the heterostructure geometry shown in FIG. \ref{fig1}(a). }
 \label{fig0}
\end{figure}

Our central finding is that the heterostructure system composed of 
an $s$-wave superconductor $\pi$-junction and ferromagnetic insulators on a topological insulator 
(as shown in Fig.\ref{fig1}(a)) 
described by (\ref{eq12})
has the index: 
\begin{equation}
\begin{split}
\mathrm{ind}\, \mathcal{H} =& \frac{1}{2} \left[ \mathrm{sign}(h_+) - \mathrm{sign}(h_-) \right] \\
&+ \mathrm{sign}(h_+) N_{x \rightarrow \infty} - \mathrm{sign}(h_-) N_{x \rightarrow -\infty}, 
\end{split}
\label{eq00}
\end{equation}
where $\mathrm{sgn}(h_{\pm})$ is the sign of the asymptotic Zeeman field $h_2(x \rightarrow \pm \infty,y)$, 
and $N_{x \rightarrow \pm \infty}$ are integer numbers which count how many times the band inversion
occurs for massive bound states at the $\pi$-junction, as the Zeeman magnetic field increases from zero to
$h_2(x \rightarrow \pm \infty,y)$. (see Fig. \ref{fig0} and discussion given at the end of this section.)
Note that in Eq.(\ref{eq00}), 
we take the origin of the $xy$ coordinate $(x,y)=(0,0)$ at the location of the point defect in Fig.\ref{fig1}(a).
It is also naturally assumed that the sign of $h_2(x \rightarrow \pm \infty,y)$ is independent of $y$.
Especially in the cases of the uniform asymptotic Zeeman field, $h(x \rightarrow \pm \infty, y) \equiv  h_{\pm}$, 
the index (\ref{eq00}) is simplified to
\begin{equation}
\begin{split}
\mathrm{ind}\, \mathcal{H} =& \frac{1}{2} \left[ \mathrm{sign}(h_+) - \mathrm{sign}(h_-) \right] \\
&+ \left[ \mathrm{sign}(h_+) \sum_{{E_{n}<|h_+|} } - \mathrm{sign}(h_-) \sum_{{E_{n}<|h_-|}} \right], 
\end{split}
\label{eq0}
\end{equation}
where $h_{+}$ ($h_{-}$) is a Zeeman field at the $\pi$-junction interface for $x>0$ ($x<0$), 
which induces mass gap of the one-dimensional gapless helical Majorana mode localized at the junction interface,  
and $E_n (>0) $ denote the absolute value of the mass gaps of the one-dimensional massive modes localized at the junction interface. 
The sum in (\ref{eq0}) is taken only for one part of the Kramer's pair. 
As mentioned before, the sum in (\ref{eq0}) represents the number of times
the band inversion occurs for the massive bound states, as $h_{\pm}$ increase from zero to finite values.
Because of particle-hole symmetry, this counting can be expressed only by $E_n>0$, as shown in Eq. (\ref{eq0}).


The index (\ref{eq0}) (or (\ref{eq00})) expresses the number of zero energy Majorana bound states in 
a point-like defect at the junction of a
chiral-symmetric superconductor (class BDI {in the Altland-Zirnbauer symmetry classes\cite{Schnyder,Kitaev2}}). 

The index (\ref{eq0}) is interpreted as the phase winding of the superconducting gap $\Delta$ around the point defect.
In the case of the $\pi$-junction with a Zeeman field as shown in the FIG. \ref{fig1}(a), 
the change of the phase of the gap function can be defined in the following way by using the 
Teo-Kane's adiabatic argument. \cite{TK,QHZ} 

Without a Zeeman field, the $\pi$-junction possesses a helical Majorana fermion localized at the junction interface (FIG. \ref{fig0}(a)).\cite{FK} 
The Zeeman field from the ferromagnetic insulator lifts the Kramer's degeneracy, and induces a mass gap of the helical Majorana fermion (FIG. \ref{fig0}(b)). 
Adiabatic deformation of the Hamiltonian without closing the energy gap enables us to introduce a nonzero imaginary part
of the superconducting gap
$\Delta_2$ at the junction interface. 
In this process, the sign of $\Delta_2$ is determined by a Zeeman field such that $\mathrm{sign} (\Delta_2) = - \mathrm{sign}(h_2) $. 
Hence, the phase shift is $- \pi \ \mathrm{sign}(h_2)$, which is described by the first term in (\ref{eq0}). 
This contribution depends only on the sign of Zeeman field and does not depend on the detail of the junction interface.  

The second term in (\ref{eq0}) is, on the other hand, a new contribution which was not discussed in previous literatures in the context of heterostructure systems 
and depends on the detail of the junction interface. 
The superconducting gap $\Delta_1$ changes its sign at the $\pi$-junction.
If the spatial variation of the magnitude of $\Delta_1$ in the vicinity of the junction is sufficiently slow, 
there exist massive bound states localized at the junction, which come in Kramer's pairs with a mass gap $|E_1|$ (FIG. \ref{fig0}(a)). 
The Zeeman field 
parallel to $y$-axis shifts the mass gaps of the Kramer's pairs by $|E_1 \pm h_2|$, respectively(FIG. \ref{fig0}(b)). 
(Here, we assume the Zeeman field is uniform. 
If not, the mass gap of the bound states depends on the detail of Zeeman field. But the qualitative nature is unaffected.)
When $|h_2|$ reaches $|E_1|$, the energy gap at the junction interface is closed (FIG. \ref{fig0}(c)), and a band inversion occurs for $h_2 < - |E_1|$. 
After band inversion, the junction interface structure 
acquires the $- 2 \pi \ \mathrm{sign}(h_2)$ phase shift in addition to the $- \pi \ \mathrm{sign}(h_2)$ phase shift, 
resulting in the total $- 3 \pi \ \mathrm{sign}(h_2)$ phase shift (FIG. \ref{fig0}(d)). 
This additional $2 \pi$ phase production arises for each massive bound state with mass gap $|E_2|, |E_3|, \cdots$, 
which describes the second term in (\ref{eq0}). 
Therefore, massive bound states at the interface give rise to additional phase winding around the point defect formed by the intersection of the $\pi$-junction interface and the ferromagnetic domain wall shown in FIG. \ref{fig1}(a).

This new contribution from the second term of (\ref{eq0})
has an important implication  for the class D heterostructure system.
As mentioned before, the class D system is characterized by the $\mathbb{Z}_2$ invariant for the Majorana zero modes,
which is exactly the parity of the index (\ref{eq0}) obtained by switching off chiral-symmetry-breaking terms.
Thus, the $\mathbb{Z}_2$ invariant of the class D heterostructure system 
may be changed by the non-topological massive bound states.
This leads to a breakdown of topological protection of Majorana vortex modes when the second term of (\ref{eq0}) is 
an odd integer.

\section{Niemi-Semenoff Index Theorem}
\label{sec3}
In this section, for the convenience of readers, we briefly review 
the Niemi-Semenoff index theorem
which is used for the derivation of our results in the following sections. 
The Niemi-Semenoff index theorem relates the number of zero energy modes in Dirac fermion systems to the geometrical structure of spatially varying mass terms.
In particular, the index is determined by the asymptotic behaviors of mass terms at open boundaries.
We consider the Dirac Hamiltonian with chiral symmetry in $d$-dimensional space with open boundaries, 
the Hamiltonian of which is given by,
\begin{equation}
\begin{split}
\mathcal{H} &= -i \Gamma_i \partial_i + Q(\bm{x}) 
=\begin{pmatrix}
0 & \mathcal{D} \\ 
\mathcal{D}^{\dag} & 0
\end{pmatrix}, \\
\end{split}
\end{equation}
with $\mathcal{D} = - i \gamma_{i} \partial_{i} + K(\bm{x})$, for the basis that the $\Gamma$-matrices are represented as 
\begin{equation}
\begin{split}
\Gamma_i = 
\begin{pmatrix}
0 & \gamma_i \\
\gamma_i^{\dag} & 0
\end{pmatrix}, \ 
\Gamma_5 = 
\begin{pmatrix}
1 & 0 \\
0 & -1
\end{pmatrix}. 
\end{split}
\label{eq2}
\end{equation}
Here, the indices $i = 1,2,\dots, d$ are those for the spatial coordinates,
the $\gamma_{i}$ matrices are constant matrices that satisfy 
$\gamma_i \gamma_j^{\dag} + \gamma_j \gamma_i^{\dag} = \gamma_i^{\dag} \gamma_j + \gamma_j^{\dag} \gamma_i= 2 \delta_{ij}$, 
and $K(\bm{x})$ includes all background fields such as electromagnetic fields and the superconducting gap. 
The index of the Hamiltonian is defined by 
$\mathrm{ind}\,\mathcal{H} := \mathrm{dim\ ker} \ \mathcal{D}^{\dag}- \mathrm{dim\ ker} \ \mathcal{D}$, 
which is the difference between the number of zero energy states of $\mathcal{H}$ with the opposite chirality. 
We assume all background fields are asymptotically independent of the normal coordinate, 
$\hat n_i(\bm{x}) \partial_i Q(x) \rightarrow 0 \ (|\bm{x}| \rightarrow \infty)$, 
where $\hat n(\bm{x})$ is a unit vector normal to an open boundary at $|\bm{x}|\rightarrow \infty$. 
It is known that $\mathrm{ind}\ \mathcal{H}$ is expressed as the sum of the volume integral of 
the chiral anomaly and the surface integral of the chiral current, \cite{Weinberg,NS1} 
\begin{equation}
\begin{split}
&\mathrm{ind} \ \mathcal{H} \\
&= \int d^d \bm{x} \ \mathrm{tr} \Braket{\bm{x}|\Gamma_5|\bm{x}} + \frac{1}{2} \oint d \hat S \ \mathrm{tr} \Braket{\bm{x} | i \hat \Gamma(\bm{x}) \Gamma_5 \mathcal{H}^{-1} | \bm{x} }, 
\end{split}
\label{eq5}
\end{equation}
where $d \hat S$ is the volume element of the boundary, 
and $\hat \Gamma(\bm{x}) := \hat{n}_i(\bm{x}) \Gamma_i$. 
The definition of the terms of r.h.s. in (\ref{eq5}) needs appropriate regularization. 
In this paper, we symbolically use the expression of r.h.s. in (\ref{eq5}). 
The first term in (\ref{eq5}) is the integrated chiral anomaly which is present only in even spatial dimensions. 
When $d=2$, it is explicitly written in terms of the background field $Q(\bm{x})$ as \cite{Weinberg}
\begin{equation}
\begin{split}
\int d^2 \bm{x} \ \mathrm{tr} \Braket{\bm{x}|\Gamma_5|\bm{x}} = - \frac{1}{4 \pi} \int d^2 \bm{x} \mathrm{tr}\  i \Gamma_5 \Gamma_i \partial_i Q(\bm{x}). 
\end{split}
\label{eq6}
\end{equation}
This formula will be used later. (see Eq.(\ref{eq14}) below)
The second term in (\ref{eq5}) is the boundary integral of the chiral current density normal to the boundary, 
and this term can be rewritten as the spectral asymmetry constructed from the real part of the eigenvalues of a certain boundary operator $\mathcal{M}$ as shown below, 
\begin{equation}
\begin{split}
&\mathrm{tr}\ \Braket{\bm{x} | i \hat \Gamma(x) \Gamma_5 \mathcal{H}^{-1} |\bm{x}} \\
&= \mathrm{tr}\ \Braket{\bm{x} | i \begin{pmatrix}
0 & \hat \gamma(\bm{x}) \\
\hat \gamma^{\dag}(\bm{x}) & 0
\end{pmatrix}
\begin{pmatrix}
1 & 0 \\
0 & -1
\end{pmatrix}
\begin{pmatrix}
0 & \mathcal{D} \\
\mathcal{D}^{\dag} & 0
\end{pmatrix}^{-1} |\bm{x}} \\ 
&= \mathrm{tr}\ \Braket{\bm{x} | \left( i \hat \gamma^{\dag}(\bm{x}) \mathcal{D} \right)^{-1} + \left[ \left( i \hat \gamma^{\dag}(\bm{x}) \mathcal{D} \right)^{-1} \right]^{\dag}|\bm{x}} .
\end{split}
\end{equation}
Here $\hat \gamma(x) = \hat{n}_i(x) \gamma_i$ are the normal components of $\gamma$ matrices, 
and we used the cyclicity of trace. 
The boundary operator $\mathcal{M}$ is defined by $i \hat \gamma^{\dag}(x) \mathcal{D} = \hat \partial + \mathcal{M}$, 
\begin{equation}
\begin{split}
\mathcal{M} = \hat \gamma^{\dag}(\bm{x}) \gamma^T_i \partial_i + i \hat \gamma^{\dag}(\bm{x}) K(\bm{x}),  \\
\end{split}
\label{eq8}
\end{equation}
where $\gamma^T_{i}(\bm{x})= \gamma_i - \hat \gamma(\bm{x}) \hat{n}_i(\bm{x})$ are the tangential components of $\gamma$ matrices, 
and $\hat \partial = \hat n_i(\bm{x}) \partial_i$ is the directional derivative normal to the boundary. 
We assume $\mathcal{H}$ does not possess zero modes at infinity, which corresponds to the absence of zero modes in $\mathcal{M}$. 
Since $\mathcal{M}$ is independent of the coordinate normal to the boundary, 
we can introduce the Fourier transformation for the normal coordinate:
\begin{equation}
\begin{split}
&\frac{1}{2} \oint d \hat S \mathrm{tr} \Braket{\bm{x} | i \hat \Gamma \Gamma_5 \mathcal{H}^{-1} | \bm{x} } \\
&= \frac{1}{2} \oint d \hat S \mathrm{tr} \Braket{\bm{x} | \frac{1}{\mathcal{M} + \hat \partial} +  \frac{1}{\mathcal{M}^{\dag} - \hat \partial} | \bm{x} } \\
&= \frac{1}{4 \pi} \int_{-\infty}^{\infty} d \hat k \oint d \hat S \mathrm{tr} \Braket{\bm{x} | \frac{1}{\mathcal{M} + i \hat k} +  \frac{1}{\mathcal{M}^{\dag} - i \hat k} | \bm{x} }.  \\
\end{split}
\label{eq9_1}
\end{equation}
Introducing the eigenmodes $\mathcal{M} \phi = \lambda \phi$ and $\mathcal{M}^{\dag} \psi = \lambda^* \psi$, 
we rewrite Eq. (\ref{eq9_1}) as 
\begin{equation}
\begin{split}
&\frac{1}{4 \pi} \int_{-\infty}^{\infty} d \hat k \int d \lambda \ \rho(\lambda) \left( \frac{1}{\lambda + i \hat k} +  \frac{1}{\lambda^* - i \hat k} \right)\\
&= \frac{1}{2} \int d \lambda \ \rho(\lambda)  \mathrm{sign} \left[ \mathrm{Re}(\lambda) \right] \\
&=: \frac{1}{2} \eta \left( \mathrm{Re} \left( \mathcal{M} \right) \right), 
\end{split}
\label{eq10}
\end{equation}
where $\rho(\lambda)$ is the spectral density of boundary operator $\mathcal{M}$. 
This term is the spectral asymmetry constructed from the real part of the eigenvalues of $\mathcal{M}$. 
Eventually, $\mathrm{ind}\,\mathcal{H}$ is written as \cite{NS1}
\begin{equation}
\begin{split}
\mathrm{ind}\, \mathcal{H} 
= \int d^d x \ \mathrm{tr} \Braket{\bm{x}|\Gamma_5|\bm{x}} + \frac{1}{2} \eta \left( \mathrm{Re} \left( \mathcal{M} \right) \right). 
\end{split}
\label{eq11}
\end{equation}
This is the Niemi-Semenoff index theorem for an open infinite space. \cite{NS1} 
The integrand of the anomaly contribution is generally the total derivative.
Hence $\mathrm{ind} \ \mathcal{H}$ depends solely on the asymptotic behavior of background fields.

\section{Majorana zero modes at a point defect in superconductor-ferromagnet insulator heterostructure systems}
\label{sec4}

In this section, 
we derive the index (\ref{eq0}) for the topological heterostructure system depicted in Fig. \ref{fig1}(a)
by applying the Niemi-Semenoff index theorem explained in the previous section.
For this purpose, we first obtain the boundary operator (\ref{eq8}) for the Hamiltonian (\ref{eq12}).
This is achieved by the following procedure.
By applying the unitary transformation,
\begin{equation}
\begin{split}
\Gamma_5 = \tau_3 \sigma_3 = \begin{pmatrix}
\sigma_3 & 0 \\
0 & -\sigma_3
\end{pmatrix}
\mapsto U \tau_3 \sigma_3 U^{\dag} = \begin{pmatrix}
1 & 0 \\
0 & -1
\end{pmatrix}
\end{split}
\end{equation}
with 
\begin{equation}
\begin{split}
U = \begin{pmatrix}
1 & & & \\
& & & 1 \\
& & 1 & \\
& 1 & &
\end{pmatrix},
\end{split}
\end{equation}
the Hamiltonian (\ref{eq12}) is represented as
\begin{widetext}
\begin{equation}
\begin{split}
\mathcal{H} \mapsto  
\begin{pmatrix}
0 & v ( \sigma_2 \partial_x - \sigma_1 \partial_y) + h_2 \sigma_2 + \Delta_1  \\
-v ( \sigma_2 \partial_x - \sigma_1 \partial_y) + h_2 \sigma_2 + \Delta_1 & 0 
\end{pmatrix}.
\end{split}
\end{equation}
\end{widetext}
In this representation, $( \gamma_1 , \gamma_2 ) = (i\sigma_2, -i \sigma_1) $, and $K(x,y)= h_2 \sigma_2 + \Delta_1$. 
Then, the boundary operator $\mathcal{M}$ defined by (\ref{eq8}) is
\begin{equation}
\begin{split}
\mathcal{M} = i v \sigma_3 \partial_T + \Delta_1 \sigma_T 
+ h_T - i \hat h \sigma_3, 
\end{split}
\label{eq13}
\end{equation}
where $\hat a = \hat n_i a_i$ and $a_T = n^T_i a_i$ are components of a vector $\bm{a} = (a_1,a_2)$
which are, respectively, normal and tangential to an open boundary at $|x|\rightarrow \infty $ or $|y|\rightarrow \infty$. 
(We note that the origin of the $xy$ coordinate $(x,y)=(0,0)$ is taken at the position of the point defect in Fig. \ref{fig1}(a).)
From (\ref{eq6}), the anomaly part of the index is  
\begin{equation}
\begin{split}
\int d^2 \bm{x} \mathrm{tr} \Braket{\bm{x}|\Gamma_5|\bm{x}}
&= \frac{1}{\pi} \int d^2 \bm{x} \epsilon_{i j} \partial_i h_j = \frac{1}{\pi} \oint d \bm{l} \cdot \bm{h}. 
\end{split}
\label{eq14}
\end{equation}
To simplify the analysis, we consider the kink structure for the gap function $\Delta_1(y) = \Delta \tanh(y / \xi)$ with $\Delta>0$ at the interface of $\pi$-junction.
As we shall see momentarily, the parameter $\xi$ in the gap function describing the width of the kink gives rise to a crucial effect on the index.
Here, it is natural to assume that
the asymptotic value of the magnetic field,
$h_2(x,y) \rightarrow h_{\pm}(y)$ as $x \rightarrow \pm \infty$, 
 have nonzero values with definite signs, $\mathrm{sign}(h_{\pm})$,
near the interface of the $\pi$-junction.
On the other hand, the magnitudes of 
$h_{\pm}(y) $ at $y \rightarrow \pm \infty$ does not affect the index of our system.
Thus, for simplicity, we assume  $h_{\pm}(y) \rightarrow 0$ for $y \rightarrow \pm \infty$.

Now we calculate the spectral asymmetry $\eta(\mathrm{Re}(\mathcal{M}))$ for the heterostructure system
depicted in Fig. \ref{fig1}(a). 
It follows from (\ref{eq13}) that the boundary operators $\mathcal{M}$ for $x \rightarrow \pm \infty$ 
and $y \rightarrow \pm \infty$
are given by,
\begin{align}
&\mathcal{M}_{x \rightarrow \infty}(y) = i v \sigma_3 \partial_y + \Delta_1(y) \sigma_2 + h_+(y), \label{boundary_1} \\
&\mathcal{M}_{x \rightarrow -\infty}(y) = -i v \sigma_3 \partial_y - \Delta_1(y) \sigma_2 - h_-(y), \label{boundary_2} \\
&\mathcal{M}_{y \rightarrow \infty}(x) = -i v \sigma_3 \partial_x - \Delta \sigma_1,  \\
&\mathcal{M}_{y \rightarrow -\infty}(x) = i v \sigma_3 \partial_x - \Delta \sigma_1 .
\end{align}
The spectral asymmetry (\ref{eq10}) is the sum of the partial spectral asymmetries of four sides. 
Each boundary operator is hermitian, 
since the Zeeman field normal to the boundary in (\ref{eq13}) vanishes. 
The spectral asymmetries for $\mathcal{M}_{y \rightarrow \pm \infty}$ are zero, 
since the eigenvalues of $\mathcal{M}_{y \rightarrow \pm \infty}$ come in pairs $\pm \lambda$ 
due to ``chiral'' symmetry $\sigma_2 \mathcal{M}_{y \rightarrow \pm \infty} \sigma_2 = -\mathcal{M}_{y \rightarrow \pm \infty}$. 
To calculate the spectral asymmetries for ${\cal M}_{x\rightarrow\pm\infty}$,
we exploit an approach developed by Lott: \cite{Lott,NS2}
Let $\mathcal{H_{\tau}}$ be a one-parameter family of Hamiltonians defined in $\tau \in [0, 1]$ which interpolate between a reference Hamiltonian ${\cal H}_0$ and ${\cal H}_1={\cal M}_{x\rightarrow\pm\infty}$. 
In the calculation of the spectral asymmetry, we choose the reference Hamiltonian ${\cal H}_0$ for which
the spectral asymmetry $\eta(\mathcal{H}_0)$ is known. 
The variance of the spectral asymmetry as a function of $\tau$ is composed of two parts: one is the 
continuum part $\eta^c_{\tau}$ raised by the change of high energy continuum energy spectrum, and 
the other is a discrete part arising from the spectral flow which changes by $\pm 2$ when a discrete eigenvalue $\lambda_n(\tau)$ crosses zero from 
negative (positive) to positive (negative) energies: $\Delta \left[ \mathrm{sign}(\lambda_n(\tau)) \right] = \pm 2$. 
Thus, we can write the spectral asymmetry in the form,
\begin{equation}
\begin{split}
\eta (\mathcal{H}) = \eta (\mathcal{H}_0) + \int d \tau \frac{d \eta^c_{\tau}}{d \tau} + 2 (\text{spectral flow}).
\end{split}
\label{eq20}
\end{equation}
Let us consider the spectral asymmetry for $x \rightarrow \infty$. In this case, the boundary operator ${\cal M}_{x\rightarrow\infty}$ is basically the Jackiw-Rebbi Hamiltonian.
Therefore, in the absence of a magnetic field, $h_+(y) \equiv 0$, $\mathcal{M}_{x \rightarrow \infty}$ possesses the Jackiw-Rebbi zero mode localized at the interface of $\pi$-junction, 
$\phi_0(y)\propto {}^t (1,-1) e^{-\int^y \Delta_1(y')/v d y'}$. 
When a magnetic field is switched on,
a finite value of $h_+(y)$ shifts the bound state energy from zero to a nonzero value with the same sign as $h_+$. 
The spectral asymmetry for this boundary operator was previously computed by Lott \cite{Lott}.
Using the reference Hamiltonian given by 
$\mathcal{H}_0 = i v \sigma_3 \partial_y + \Delta_1(y) \sigma_2 - \delta \mathrm{sign}(h_+) \sigma_1$ 
with a small positive constant $\delta$, 
we obtain, 
\begin{equation}
\begin{split}
&\frac{1}{2} \eta(\mathcal{M}_{x \rightarrow \infty}) \\
&=  \frac{1}{2} \mathrm{sign} (h_+) - \frac{1}{\pi} \int d y h_+(y) + (\text{spectral flow}).
\end{split}
\label{eq21}
\end{equation}
The first term conforms to the fermion fractionalization in the Jackiw-Rebbi system.\cite{JR,GW}
The second term is the volume part of the variation of $\eta_{\tau}$, which cancels out the anomaly contribution of the index (\ref{eq14}). 
The third term is the spectral flow contribution from $\mathcal{H}_0$ to $\mathcal{M}_{x \rightarrow \infty}$ which depends on the structure of $\Delta_1(y)$ and $h_+(y)$. 
The spectral flow stems from the bound states localized at the interface of the $\pi$-junction. 
Since the Lott's derivation of (\ref{eq21}) in ref.\onlinecite{Lott}
is highly technical, we give a more elementary derivation of (\ref{eq21}) in Appendix, which we believe is useful for readers. 
If the finite value region of $h_+(y)$ is much longer than $\xi$, $h_+(y)$ is approximated as a constant chemical potential, $h_+(y) \rightarrow h_+$. 
In this case, the $y$-dependent part of $\mathcal{M}_{x \rightarrow \infty}$, i.e., $i v \sigma_3 \partial_y + \Delta \tanh(y/\xi)$ 
is exactly solvable \cite{SK,Takayama} 
The eigenvalues of this Hamiltonian are $E_0 = 0$, 
\begin{equation}
\begin{split}
E_{n,\pm} = \pm \Delta \sqrt{\frac{n}{\nu} \left( 2-\frac{n}{\nu}\right)} , (n = 1,2,\dots,<\nu),  \\
\end{split}
\label{eq22}
\end{equation}
and $E_{p,\pm} = \pm \sqrt{v^2 p^2 + \Delta^2} ,\  (p \in \mathbb{R})$, where $\nu = \xi \Delta/v = \xi/\xi_c$ is a ratio of $\xi$ to the coherence length of the superconducting state : $\xi_c = v/\Delta$. 
$E_0$ and $E_{n,\pm}$ are the energy of the bound states localized at the interface of $\pi$-junction. 
The Jackiw-Rebbi zero energy bound state exists for arbitrary $\xi$, 
while the massive bound states exist only when $\xi > \xi_c$. 
\begin{figure}[!]
 \begin{center}
  \includegraphics[width=0.9\linewidth, trim=0.5cm 2.0cm 0.5cm 2.3cm]{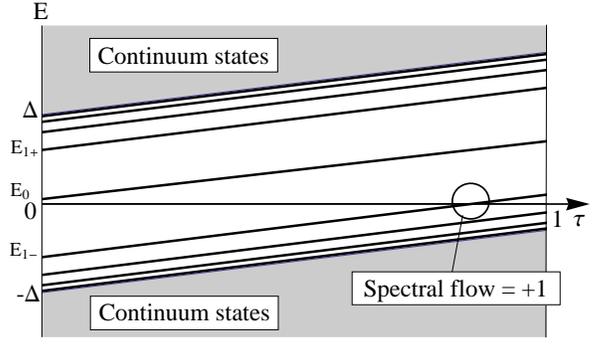}
 \end{center}
 \caption{The $\tau$-dependent energy spectrum of a one parameter family of Hamiltonians $\mathcal{H}_{\tau}$: 
 $\mathcal{H}_{\tau} = (1-\tau) \mathcal{H}_0 + \tau \mathcal{H}$. 
 This figure shows the case of $\xi = 5 \xi_c$, $h_+ = 0.7 \Delta> 0$, and $\delta = 0.05 \Delta$. 
 }
 \label{fig2}
\end{figure}
$h_+$ induces constant shifts to eigenvalues (\ref{eq22}), 
and hence, the spectral flow is given by the bound states that cross zero between 
$E_{n\pm} $ and $E_{n\pm} + h_+ $ as shown in Fig. \ref{fig2}: 
\begin{equation}
\begin{split}
(\text{spectral flow}) = \mathrm{sign}(h_+) \sum_{{E_{n}<|h_+|} }, 
\end{split}
\label{eq23}
\end{equation}
where 
$E_n=E_{n,+}$.
Note that due to the term $- \delta \mathrm{sign} (h_+) \sigma_1$ in the reference Hamiltonian $\mathcal{H}_0$, 
the spectral flow from the Jackiw-Rebbi zero energy bound state is excluded in the sum (\ref{eq23}). 
In a similar way, the spectral asymmetry of $\mathcal{M}_{x \rightarrow -\infty}$, $\frac{1}{2} \eta(\mathcal{M}_{x \rightarrow -\infty})$ can be calculated as 
\begin{equation}
\begin{split}
- \frac{1}{2} \mathrm{sign} (h_-) +\frac{1}{\pi} \int d y h_-(y) - \mathrm{sign}(h_-) \sum_{{E_{n}<|h_-|}}.
\end{split}
\label{eq24}
\end{equation}
Using Eqs. (\ref{eq11}), (\ref{eq14}), (\ref{eq21}), (\ref{eq23}) and (\ref{eq24}) together, 
we arrive at the formula (\ref{eq0}). 

So far we have calculated the index for (\ref{eq12}). 
The $\mathbb{Z}_2$ index $N$ for the vortex zero modes in class D heterostructure system (\ref{eq11.5}) is, as we have mentioned, given by $N = \mathrm{ind} \, \mathcal{H} \ (\mathrm{mod} \ 2)$. 
Remarkably, the second term in (\ref{eq0}), which is basically the contribution from non-topological bound states of  
the Jackiw-Rebbi Hamiltonian, can affect the $\mathbb{Z}_2$ index. 
This indicates that the existence of the non-Abelian vortex zero modes depends on the kink-structure of the gap function parametrized by $\xi$.
Actually, in the case that the second term of (\ref{eq0}) is equal to an odd integer, which can indeed occur when $\xi > \xi_c$ (i.e. $\nu>1$),
the $\mathbb{Z}_2$ invariant for Majorana zero modes is changed,
resulting in the breakdown of the topological protection for the $\mathbb{Z}_2$ Majorana modes. 
We now discuss the condition for which $\xi > \xi_c$ is realized.
Actually, to determine $\xi$ precisely, we need to solve the Bogoliubov-de-Gennes equation
for proper boundary conditions, which is out of the scope of this paper.
Instead of presenting such precise analysis, we here give a qualitative argument.
For the superconductor-ferromagnetic insulator junction as depicted in Fig. 1 (a), the gap function at the junction
is reduced by magnetic scattering at the interface between the superconductor and the ferromagnet.\cite{toku}
On the other hand, the dimension of the ferromagnet insulator along  the $y$-axis denoted as $L_y$ 
plays the role of the characteristic length scale for the spatial variation of the exchange field along the $y$-axis.
Thus, when $L_y$ is sufficiently larger than the coherence length $\xi_c$,
 we can neglect the spatial variation of the superconducting gap
raised by magnetic scattering near the interface, and hence it is expected that $\xi < \xi_c$ is satisfied, ensuring the topological protection of Majorana modes.
However, when $L_y$ is comparable to $\xi_c$,  the spatially inhomogeneous
reduction of the superconducting gap due to magnetic scattering
crucially affects the magnitude of the parameter $\xi$.
In particular, when $L_y$ is slightly larger than $\xi$,
it may be possible that $\xi > \xi_c$ is realized, which leads to
the above-mentioned mechanism of the breakdown of $\mathbb{Z}_2$ nontriviality.


\section{Zero modes in line defects of topological insulator-ferromagnet insulator heterostructure junctions}
\label{sec5}
The index theorem (\ref{eq0}) is also applicable to 
a topological insulator-ferromagnetic insulator tri-junction system, the setup of which is depicted in Fig. \ref{fig1} (b). 
The two orbital effective Dirac model for this system is written as \cite{QHZ}
\begin{equation}
\begin{split}
\mathcal{H} = v k_z \mu_1 \sigma_3 -i v \mu_1 \sigma_j \partial_j + m(y) \mu_3 + h_2(x,y) \sigma_2 -\mu ,
\end{split}
\end{equation} 
where $j = 1,2 $, $\bm{\mu}= (\mu_1, \mu_2, \mu_3)$ and $\bm{\sigma} = (\sigma_1,\sigma_2, \sigma_3)$ are the Pauli matrices for the orbital and the spin spaces, respectively, 
$v$ is the velocity of the Dirac fermion, $m$ is the mass gap whose sign determines whether the system is in a topological $(m<0)$ or trivial $(m>0)$ phase, $\bm{h} \cdot \bm{\sigma}$ is a Zeeman term, and $\mu$ is the chemical potential. 
We have assumed the translational invariance along the $z$-direction. 
This system belongs to the class A, and the chiral gapless modes localized at line defects are classified as $\mathbb{Z}$.\cite{TK}
As in the case of the superconductor-ferromagnet insulator junction, 
we, first, neglect  chiral-symmetry breaking terms, putting $\mu=0$:
\begin{equation}
\begin{split}
\mathcal{H} &=v k_z \mu_1 \sigma_3 -i v \mu_1 \sigma_i \partial_i + m(y) \mu_3 + h_2(x,y) \sigma_2 \\
&=: v k_z \mu_1 \sigma_3 + \tilde{ \mathcal{H}}(x,y) ,
\end{split}
\end{equation}
Because of chiral symmetry $\{ \mu_1 \sigma_3 , \tilde{ \mathcal{H}}(x,y) \} = 0$, 
the chiral zero bound states of $\tilde{ \mathcal{H}}(x,y)$ with chirality $\pm$ correspond to the chiral gapless modes with the energy dispersion 
$\pm v k_z$ with chirality $\pm$. 
Since $\tilde{ \mathcal{H}}(x,y)$ is of the same form as (\ref{eq12}), $\mathrm{ind} \,\tilde{ \mathcal{H}}$ is given by (\ref{eq0}), 
but in this case, $E_n(>0)$ is the mass gap of the two-dimensional massive bound states localized at the surface of topological insulator. 
The first contribution in (\ref{eq0}) agrees with the winding of the Axion vortex.\cite{QHZ} 
The second contribution in (\ref{eq0}) correspond to the non-topological integer part of Axion field which depend on the microscopic structure of the interface between 
the topological insulator and the trivial insulator. 
The chiral gapless mode cannot be massive since backward scatterings are suppressed. 
This index for chiral gapless modes in class A survives against any perturbations.

\section{conclusion and discussions}
\label{sec6}
Here, we remark a topological property for the index of heterostructure systems. 
The index (\ref{eq0}) is stable against the continuous change of Hamiltonian $\mathcal{H}$ 
unless the boundary operator $\mathcal{M}$ have zero modes, i.e., unless the Hamiltonian $\mathcal{H}$ have no gapless modes at infinity. 
In this sense, the index (\ref{eq0}) is topologically protected by the energy gap at the boundary. 
This feature is similar to the topological order in bulk systems protected by a bulk energy gap. 

In conclusion, we have shown that 
the number of Majorana bound states in the $\pi$-junction-ferromagnet heterostructure systems is 
affected 
by massive bound state localized at the interface, 
which has an important implication for topological protection of zero modes in class D systems. 



This work was supported in part by the Grant-in-Aids for
Scientific Research from MEXT of Japan [Grants No. 23102714, No. 23540406, and No. 23103502(Innovation Areas ``Topological Quantum Phenomena'')], 
and from the Japan Society for the Promotion of Science (Grant No. 21540378).

\appendix
\label{app}
\section{Derivation of (\ref{eq21})}
In this appendix, we derive Eq.(\ref{eq21}) by using
the Niemi-Semenoff formula for the fermion number fractionalization, \cite{NS2}
which is an alternative expression of the Niemi-Semenoff index theorem. 

\subsection{Niemi-Semenoff formula of fermion number fractionalization}
We introduce an extended Hamiltonian $\tilde{\mathcal{H}}(\tau,\bm{x})$ defined by
\begin{equation}
\begin{split}
\tilde{\mathcal{H}}(\tau,\bm{x})
&= -i \sum_{i=0}^d \Gamma_i \partial_i + Q(\tau,\bm{x}) \\
&=\begin{pmatrix}
0 & \partial_{\tau} + H(\tau,\bm{x}) \\
-\partial_{\tau} + H(\tau,\bm{x}) & 0
\end{pmatrix}, \\
\end{split}
\end{equation}
where $\partial_0 = \partial_{\tau}$, and
$H(\tau,\bm{x}) = -i \sum_{i = 1}^d \gamma_i \partial_i + K(\tau,\bm{x})$ 
satisfies $H(\tau  \rightarrow \infty,\bm{x}) = H_1(\bm{x})$, and $H(\tau  \rightarrow -\infty,\bm{x}) = H_0(\bm{x})$
with $H_1(\bm{x})$ the target Hamiltonian for which we want to calculate the spectral asymmetry,
and $H_0(\bm{x})$ a reference Hamiltonian.
$H(\tau,\bm{x}) $ 
interpolates between $H_1(\bm{x})$ and $H_0(\bm{x})$ as a function of the auxiliary parameter $\tau$. 
Also, a new gamma matrices $\gamma_0 = i$ is introduced. 
From the Niemi-Semenoff index theorem (\ref{eq11}), the index of $\tilde{\mathcal{H}}$ is given by 
\begin{equation}
\begin{split}
\mathrm{ind} \ \tilde{\mathcal{H}} = \int d \tau d^d \bm{x} \ \mathrm{tr} \Braket{\tau,\bm{x}|\Gamma_5|\tau,\bm{x}} + \frac{1}{2} \eta \left( \mathrm{Re}(\mathcal{M}) \right). 
\end{split}
\label{aeq1}
\end{equation}
Note that there are three boundary operators which contribute to 
the spectral asymmetry $\eta \left( \mathrm{Re}(\mathcal{M}) \right)$; i.e.
$\mathcal{M_{\tau \rightarrow \infty}} $, $\mathcal{M_{\tau \rightarrow -\infty}} $, and
$\mathcal{M}_{|\bm{x}| \rightarrow \infty}$.
The first two boundary operators are, respectively, related to $H_1(\bm{x})$ and $H_0(\bm{x})$,
\begin{equation}
\begin{split}
&\mathcal{M_{\tau \rightarrow \infty}} = \hat \gamma^{\dag} \gamma_i \partial_i + i \hat \gamma^{\dag} K(\tau  \rightarrow \infty,\bm{x}) = H_1(\bm{x}),  \\
&\mathcal{M_{\tau \rightarrow -\infty}} = \hat \gamma^{\dag} \gamma_i \partial_i + i \hat \gamma^{\dag} K(\tau  \rightarrow -\infty,\bm{x}) = - H_0(\bm{x}),  \\
\end{split}
\end{equation}
where, $\hat \gamma(\tau \rightarrow \pm \infty) = \pm \gamma_0$. 
Then, from Eq.(\ref{aeq1}), we have\cite{NS2}
\begin{equation}
\begin{split}
&\frac{1}{2} \eta \left( H_1 \right) 
= \frac{1}{2} \eta \left( H_0 \right) + \mathrm{ind} \ \tilde{\mathcal{H}} \\
&\ \ - \int d \tau d^d \bm{x} \ \mathrm{tr} \Braket{\tau,\bm{x}|\Gamma_5|\tau,\bm{x}} - \frac{1}{2} \eta \left( \mathrm{Re}(\mathcal{M}_{|\bm{x}| \rightarrow \infty}) \right), 
\end{split}
\label{a4}
\end{equation}
where we have used $\eta(-H) = -\eta(H)$. 
Generally, the change of the spectral asymmetry is divided into its continuous part and discontinuous part, 
\begin{equation}
\begin{split}
\eta(H_1) - \eta(H_0) = \int d\tau \frac{d \eta^c_{\tau}}{d \tau} + 2 (\text{spectral flow}) . 
\end{split}
\label{a5}
\end{equation}
Comparing (\ref{a4}) and (\ref{a5}), we find,
\begin{equation}
\begin{split}
&(\text{spectral flow}) = \mathrm{ind}\  \tilde{\mathcal{H}}, \\
\end{split}
\label{a6}
\end{equation}
Then the spectral asymmetry of $H_1$ is given by 
\begin{equation}
\begin{split}
&\frac{1}{2} \eta \left( H_1 \right) 
= \frac{1}{2} \eta \left( H_0 \right) + (\text{spectral flow}) \\
&\ \ - \int d \tau d^d \bm{x} \ \mathrm{tr} \Braket{\tau,\bm{x}|\Gamma_5|\tau,\bm{x}} - \frac{1}{2} \eta \left( \mathrm{Re}(\mathcal{M}_{|\bm{x}| \rightarrow \infty}) \right). 
\end{split}
\label{a61}
\end{equation}

\subsection{The spectral asymmetry of the Hamiltonian (\ref{boundary_1}) and (\ref{boundary_2})}
We apply (\ref{a61}) to the boundary operator (\ref{boundary_1}), 
\begin{equation}
\begin{split}
H_1(y) = i v \sigma_3 \partial_y + \Delta_1(y) \sigma_2 + h_+(y),  \\
\end{split}
\end{equation}
and the reference Hamiltonian, 
\begin{equation}
\begin{split}
H_0(y) = i v \sigma_3 \partial_y + \Delta_1(y) \sigma_2 - \delta \sigma_1, 
\end{split}
\end{equation}
where $\delta$ is a small constant which is introduced to suppress zero energy modes of $H_0(y)$.
We introduce a Hamiltonian that interpolates $H_1(y)$ and $H_0(y)$, 
\begin{equation}
\begin{split}
H(\tau,y) = i v \sigma_3 \partial_y + \Delta_1(y) \sigma_2 + \Delta_2(\tau,y) \sigma_1 + h_2(\tau,y), 
\end{split}
\end{equation}
where $\Delta_2(\tau \rightarrow \infty, y) = 0$, $\Delta_2(\tau \rightarrow -\infty, y) = -\delta$, $h_2(\tau \rightarrow \infty, y) = h_+(y)$, and $h_2(\tau \rightarrow -\infty, y) = 0$. 
We assume $\Delta(\tau,y)$ and $h_2(\tau,y)$ form a single kink structure along the $\tau$ direction. 
The boundary operator at $y \rightarrow \infty$ is
\begin{equation}
\begin{split}
&\mathcal{M}_{y \rightarrow \infty} \\
&= \hat \gamma^{\dag} \gamma_i \partial_i + i \hat \gamma^{\dag} K(\tau,y \rightarrow \infty) \\
&= -i \sigma_3 \partial_{\tau} - i \sigma_3 K(\tau,y \rightarrow \infty) \\
&= -i \sigma_3 \partial_{\tau} - \Delta_1(\infty) \sigma_1 + \Delta_2(\tau, \infty)\sigma_2 - i h_2(\tau, \infty) \sigma_3,
\end{split}
\label{a11}
\end{equation}
where $\hat \gamma ( y \rightarrow \infty)= \gamma_1 = - \sigma_3$. 
The fourth term in (\ref{a11}) does not contribute to the spectral asymmetry, since $- i h_2(\tau, \infty) \sigma_3$ is anti-hermite. 
Thus, the spectral asymmetry $\eta \left( \mathrm{Re} \left( \mathcal{M}_{y \rightarrow \infty} \right) \right)$ 
arises from the Jackiw-Rebbi Hamiltonian, 
$-i \sigma_3 \partial_{\tau} - \Delta_1(\infty) \sigma_1 + \Delta_2(\tau, \infty)\sigma_2$.
The spectral asymmetry of this Hamiltonian is well known, \cite{NS2} and 
equal to the phase winding of $-\Delta_1(\infty) + i \Delta_2(\tau, \infty)$ raised by changing $\tau$ from $-\infty$ to $ \infty$ :
\begin{equation}
\begin{split}
\frac{1}{2} \eta \left( \mathrm{Re} \left( \mathcal{M}_{y \rightarrow \infty} \right) \right) = -\frac{1}{2 \pi} \mathrm{Arctan}\left( \frac{\delta}{\Delta_1(\infty)} \right) ,  
\end{split}
\label{a12}
\end{equation}
where $\mathrm{Arctan}$ has principal values, $-\frac{\pi}{2} < \mathrm{Arctan} < \frac{\pi}{2}$. 
Similarly, the spectral asymmetry of the boundary operator at $y \rightarrow -\infty$ is computed as,
\begin{equation}
\begin{split}
\frac{1}{2} \eta \left( \mathrm{Re} \left( \mathcal{M}_{y \rightarrow -\infty} \right) \right) = \frac{1}{2 \pi} \mathrm{Arctan} \left( \frac{\delta}{\Delta_1(-\infty)} \right) . 
\end{split}
\label{a13}
\end{equation}
On the other hand, the spectral asymmetry of the reference Hamiltonian $H_0$ is 
\begin{equation}
\begin{split}
\frac{1}{2} \eta \left( H_0 \right) = \frac{1}{2 \pi} \left[ \mathrm{Arctan} \left( \frac{\Delta_1(\infty)}{\delta} \right) - \mathrm{Arctan} \left( \frac{\Delta_1(-\infty)}{\delta} \right) \right]. 
\end{split}
\label{a14}
\end{equation}
Summing up the boundary contributions (\ref{a12}), (\ref{a13}) and the contribution from the reference Hamiltonian (\ref{a14}), 
we obtain,
\begin{equation}
\begin{split}
&\frac{1}{2} \eta \left( H_0 \right) - \frac{1}{2} \eta \left( \mathrm{Re}(\mathcal{M}_{y \rightarrow \infty}) \right) - \frac{1}{2} \eta \left( \mathrm{Re}(\mathcal{M}_{y \rightarrow -\infty}) \right) \\
&= \frac{1}{4} \left[ \mathrm{sgn}\left( \frac{\Delta_1(\infty)}{\delta} \right) - \mathrm{sgn}\left( \frac{\Delta_1(-\infty)}{\delta} \right) \right], 
\end{split}
\label{a15}
\end{equation}
where we have used $\mathrm{Arctan}(x) + \mathrm{Arctan}(x^{-1}) = \frac{\pi}{2} \mathrm{sgn}(x)$. 
If $\Delta_1(\infty) > 0$, $\Delta_1(-\infty)<0$, and $\mathrm{sgn}(\delta) = \mathrm{sgn}(h_+)$ ($\mathrm{sgn}(h_+)$ is the sign of the Zeeman field $h_+(y)$), then
(\ref{a15}) is written as, 
\begin{equation}
\begin{split}
&\frac{1}{2} \eta \left( H_0 \right) - \frac{1}{2} \eta \left( \mathrm{Re}(\mathcal{M}_{y \rightarrow \infty}) \right) - \frac{1}{2} \eta \left( \mathrm{Re}(\mathcal{M}_{y \rightarrow -\infty}) \right) \\
&= \frac{1}{2} \mathrm{sgn}(h_+). 
\end{split}
\label{a16}
\end{equation}
Next, the anomaly part in (\ref{a61}) is calculated from (\ref{eq6}),  
\begin{equation}
\begin{split}
\int d \tau d y \ \mathrm{tr} \Braket{\tau,y|\Gamma_5|\tau,y} 
&= -\frac{1}{4 \pi} \int d\tau dy \mathrm{tr} \left[ i \Gamma_5 \Gamma_i \partial_i Q(\tau,y) \right] \\
&= -\frac{1}{4 \pi} \int d\tau dy \left( -4 \partial_{\tau} h_2(\tau,y) \right) \\
&= \frac{1}{\pi} \int dy h_+(y). 
\end{split}
\label{a17}
\end{equation}
From (\ref{a61}), (\ref{a16}) and (\ref{a17}), we obtain (\ref{eq21}).

\end{document}